\documentclass[a4paper,11pt]{article}
\usepackage[english]{babel}
\usepackage[T1]{fontenc}
\usepackage{textcomp}
\usepackage{epsfig}
\usepackage{latexsym}
\usepackage{graphicx}
\usepackage{amsfonts,amssymb,bm}
\usepackage{color}
\newtheorem{proposition}{Proposition}
\newtheorem{lemma}{Lemma}

\newcommand{\D}{{\rm d}}
\newcommand{\sign}{{\rm sign\,}}

\begin{document}
\title{An extension of Poincaré group abiding arbitrary acceleration}
\author{Josep Llosa\\
\small Departament de Física Fonamental, Universitat de Barcelona, Spain}

\maketitle

\begin{abstract}

\noindent
The class of accelerated reference frames has been studied, on the basis of Fermi-Walker coordinates. The infinitesimal transformations connecting two of these frames has been obtained, and also their commutation relations. The outcome is an infinite dimensional extension of the Poincaré algebra. Although this extension turns out to be Abelian, and hence trivial, it is noteworthy that, contrarily to what happens with Lorentz boosts, acceleration
boost generators commute with each other and with translation generators.\\[2ex]
\end{abstract}

\section{Introduction\label{S0}}
According to the principle of relativity of Galileo, the laws of [Newtonian] mechanics hold in all inertial reference frames and are covariant under Galilei's transformations. Any couple of these reference frames are in uniform rectilinear motion with respect to each other. However, this principle of relativity can be extended to arbitrary Newtonian frames, that are mutually related by coordinate transformations like
$$  x^{\prime i} = R^i_{\,j}(t)  x^j + s^i(t)\,, \qquad \qquad t^\prime = t + t_0 \,,  $$
where $\, R^i_{\,j}(t)$ is an orthogonal matrix and $s^i(t)$ arbitrary functions of time. Indeed, the laws of Newtonian mechanics are preserved by these transformations, provided that the necessary inertial force fields ---dragging, Coriolis, centrifugal, \ldots--- are also included.

In the special theory of relativity the laws of physics hold in all Lorentzian reference frames, the relative motion of any couple of these frames is rectilinear and uniform, and the coordinates in any pair of these frames are connected by a Poincaré transformation. Furthermore, every Lorentzian frame of reference is based on an outfit of synchronized clocks which are stationary at every point in a Euclidian reference space.

In his attempt to set up a theory of gravitation consistent with his theory of relativity, Einstein initially adressed the generalization of relativity to accelerated motions \cite{Einstein1907}, but he soon abandoned it in favour of the principle of general covariance which obviously allows for a much larger invariance group than Poincaré group, namely spacetime diffeomorphisms. However, as soon was pointed out by Kretschmann \cite{Kretschmann1917}, \cite{Fock} ``since any theory, whatever its physical content, can be rewritten in a generally covariant form, the group of general coordinate transformations is physically irrelevant'' \cite{Antoci2009}. More recently other authors have insisted in the convenience of restricting genral covariance \cite{Ellis1995}, \cite{Zalaletdinov1996}.

Moreover, in Kretschmann's view, special relativity is the one with the relativity postulate of largest content; indeed, its isometry group is a ten-parameter group, which is the largest group of isometry in four dimensions, whereas for generic spacetimes in general relativity the isometry group reduces to the identity.

An alternative way to extend the class of Lorentzian frames could consist of requiring the reference space to be rigid (in the sense of Born), i. e. that the infinitesimal radar distance \cite{Born1909}, \cite{Landau} keeps constant. But, as it was clear very soon \cite{Herglotz}, even in Minkowski spacetime the only permited rigid motions are: (i) rectilinear uniform motions, (ii) uniform rotational motions around a fixed axis and (iii) arbitrary accelerated motions without rotation. It was clear that arbitrary rotational motions posed a genuine obstruction in two respects: the impossibility of synchronizing stationary clocks along a space path surrounding the rotation axis and in that the space geometry associated to the infinitesimal radar distance is neither Euclidean nor rigid \cite{Ehrenfest}.

To avoid this ``no go'' we shall here restrict to non-rotational motions with arbitrary acceleration in Minkowski spacetime and 
use Fermi-Walker reference frames to embody accelerated systems of reference, for they are often seen as the natural general-relativistic generalization of inertial Cartesian coordinates \cite{Bini2015}, \cite{Marzlin1993}. We shall then prove that these are the only synchronous frames having a flat space. 
We shall then characterize the transformations connecting two Fermi-Walker coordiante systems as {\em generalised isometries } \cite{Bel93} of the spacetime metric and, by solving the corresponding generalised Killing equation, we shall obtain an infinite dimensional extension of Poincaré algebra which includes acceleration and may be taken as the mathematical counterpart of an extension of the principle of special relativity.

\section{Fermi-Walker coordinates  \label{S1}}
Let  $\,z^\mu(\tau) \,$ be a proper time parametrized timelike worldline in ordinary Minkowski spacetime, which we shall take as the {\em space origin} (we take $c=1\,,$) $\mu=1 \ldots 4\,$, $u^\mu = \dot{z}^\mu(\tau)\,$ the unit velocity vector and $a^\mu = \ddot{z}(\tau) \,$ the proper acceleration. (We refer to Lorentzian components unless the contrary is explicitely stated.)

A 4-vector $\,w^\mu(\tau)\,$ is said Fermi-Walker (FW) transported \cite{Synge1} along $z^\mu(\tau)\,$ if 
\begin{equation}  \label{E1}
\frac{\D w^\mu}{ \D \tau} = \left(u^\mu a_\nu - u_\nu a^\mu \right)\, w^\nu
\end{equation}

If $\,z^\mu(\tau) \,$ is a straight line ($a^\mu=0 \,$ and uniform rectilinear motion), Fermi-Walker transport coincides with parallel transport and the equation above reduces to $w^\mu = $constant.

It is obvious that $u^\mu(\tau) \,$ is FW transported along $\,z^\mu(\tau) \,$ whereas, generally, it is not parallel transported. Thus FW transport is the minimal modification of parallel transport such that the proper velocity 4-vector is FW transported. On its turn, proper acceleration $a^\mu(\tau) \,$ is FW transported only if worldline $z^\mu(\tau) \,$ is contained in a plane, i. e. one-directional motion.

Let us now consider an orthonormal tetrad $e^\mu_{(\alpha)}(\tau) \,$, $\alpha = 1 \ldots 4\,$, which is FW transported along $z^\mu(\tau) \,$ and such that $e^\mu_{(4)} = u^\mu \,$. From the transport law (\ref{E1}) it follows that:
\begin{equation}  \label{E2}
\frac{\D e^\mu_{(4)}}{ \D \tau} = \sum_{i=1}^3 a_i(\tau)\, e^\mu_{(i)} \,, \qquad \qquad 
\frac{\D e^\mu_{(j)}}{ \D \tau} =  a_j(\tau)\, e^\mu_{(4)}  \,, \qquad j=1 \ldots 3
\end{equation}
where
\begin{equation}  \label{E2a}
a_i(\tau) = a_\nu(\tau) \,e^\nu_{(i)}(\tau) \,, \qquad \qquad   a^\mu(\tau) = \sum_{i=1}^3 a_i(\tau)\,e^\mu_{(i)}(\tau)
\end{equation}
That is
\begin{equation}  \label{E2b}
\frac{\D e^\mu_{(\alpha)}}{ \D \tau} = \sum_{\beta=1}^4 W^\beta_{\;\;\alpha}\, e^\mu_{(\beta)} \,, \qquad \qquad 
W^\beta_{\;\;\alpha} = \left( \begin{array}{c|c}
                     0_3 & a_i \\  \hline   a_j & 0 
                     \end{array}  \right)
\end{equation}
$0_3\,$ being the null square $\, 3 \times 3\,$ matrix.
 
For a given point in spacetime with Lorentzian coordinates $\,x^\mu\,$, the Fermi-Walker coordinates \cite{Synge2}, \cite{Gourgoulon} with space origin on $z^\mu(\tau)$ are:
\begin{description}
\item[The time] $\tau(x^\nu)$,  given as an implicit function by
\begin{equation}  \label{E3}
  \left[x^\mu - z^\mu(\tau)\right]\,u_\mu(\tau) = 0
\end{equation}
\item[The space coordinates], defined by
\begin{equation}  \label{E4}
 \xi^i = \left[x_\mu - z_\mu(\tau(x))\right]\,e^\mu_{(i)}(\tau(x))
\end{equation}
\end{description}

By differentiating these two equations, it easily follows that
\begin{equation}  \label{E5}
 u_\mu\,\D x^\mu = - \left[1 + \vec{\xi}\cdot\vec{a}(\tau)\right]\, \D\tau \,, \qquad \qquad  \D\xi^i =  e^\mu_{(i)}\,\D x^\mu 
\end{equation}
where, for the sake of abbreviation, the ordinary vector notation in three dimensions has been adopted, namely $\,\vec{\xi}\cdot\vec{a} = \sum_{i=1}^3 \xi^i a_ i \,$.

As $e^\mu_{(\alpha)}\,$ is an orthonormal base, $\quad \D x^\mu = u^\mu\,\left[1 +\vec{\xi}\cdot\vec{a}(\tau)\right]\, \D\tau + \sum_{i=1}^3 e^\mu_{(i)}\, \D\xi^i \,$ and the invariant interval in FW coordinates is
\begin{equation}  \label{E6}
\D s^2 = \D\vec{\xi}^2 - \left[1 + \vec{\xi}\cdot\vec{a}(\tau) \right]^2 \D\tau^2
\end{equation}

\subsection{The Fermi-Walker reference frame associated to $z^\mu(\tau)$  \label{S1.1} }
As any FW transported tetrad $\,e^\mu_{(\alpha)}\,$ is a solution of the linear ordinary differential system (\ref{E1}), which only depends on the origin line through $u^\mu$ and $a^\mu$, we shall have that
\begin{equation}  \label{E7}
e^\mu_{(\alpha)}(\tau) = \Lambda^\mu_{\;\;\nu}(\tau)\, e^\nu_{(\alpha)}(0)
\end{equation}
where $\,\Lambda^\mu_{\;\;\nu}(\tau)\,$ is a solution of the differential system
\begin{equation}  \label{E6b}
\frac{\D \Lambda^\mu_{\;\;\nu}}{ \D \tau} = \left(u^\mu a_\alpha - u_\alpha a^\mu \right)\, \Lambda^\alpha_\nu \,, \qquad \qquad 
\Lambda^\mu_{\;\;\nu}(0) = \delta^\mu_\nu 
\end{equation}

Therefore two tetrads, $\,e^\mu_{(\alpha)}\,$ and $\,e^{\prime \mu}_{(\alpha)}\,$, which are FW transported along the same worldline can only differ in their initial values and as, besides $\,e^\mu_{(4)} = e^{\prime \mu}_{(4)}= u^\mu\,$, these initial values are connected by a space rotation 
$$ e^{\prime \mu}_{(\alpha)}(0) = \sum_{\beta=1}^4 e^\mu_{(\beta)}(0)\,R^\beta_{\;\;\alpha} \,, \qquad {\rm with} \qquad 
R^\beta_{\;\;4} = R^4_{\;\;\beta} =\delta^\beta_4  \,, \qquad $$
and $\,\left(R^i_{\;\;j}\right)_{i,j=1\ldots 3}$ being a constant orthogonal matrix. Combining the latter with (\ref{E7}) we have that
\begin{equation}  \label{E10}
e^{\prime \mu}_{(\alpha)}(\tau) = \sum_{\beta=1}^4 e^\mu_{(\beta)}(\tau)\,R^\beta_{\;\;\alpha} 
\end{equation}

Hence all FW transported tetrads along a given worldline are the same apart from an initial space rotation and, according to the definition (\ref{E3}-\ref{E4}) the Fermi-Walker coordinates based on any of these tetrads will differ at most in a constant rotation:
$$ \tau^\prime = \tau \,, \qquad \qquad \xi^i = \sum_{j=1}^3 R^i_{\;\; j} \xi^{\prime j}  $$

Given a FW coordinate system based on $\,z^\mu(\tau)\,$, the worldlines $\xi^i =$constant, $\tau \in \mathbb{R}$ are the ``history'' of a place in the reference space associated to the FW coordinates and, as commented above, in any other FW coordinated system based on   $\,z^\mu(\tau)\,$ we shall still have that
$$ \xi^{\prime i} = \sum_{j=1}^3 \,\left(R^{-1}\right)^i_{\;\;j} \xi^j = {\rm constant}  $$

According to the invariant interval (\ref{E6}), the infinitesimal radar distance \cite{Landau},\cite{Born1909} between two close places $\xi^j$ and $\xi^j +\D\xi^j$ in the Fermi-Walker reference space is $\, \D l^2 = \D \vec\xi^2 \,$. The reference space is therefore rigid and flat, and Fermi-Walker coordinates are cartesian coordinates in this space. We shall thus profit of the ordinary vector notation $\vec\xi =(\xi^1,\xi^2,\xi^3)$.

In Lorentzian coordinates the worldline of the place $\vec\xi$ reads
\begin{equation}  \label{E6a}
\varphi^\mu(\tau; \vec\xi) = z^\mu(\tau) + \sum_{i=1}^3\xi^i\,e^\mu_{(i)}(\tau) \,, \qquad \qquad \tau \in\mathbb{R}  
\end{equation}
The proper time rate at this place is 
$$  \D\hat\tau =  \left[1 + \vec\xi\cdot\vec{a}(\tau) \right]\,\D\tau \,;$$
$\hat\tau$ is the time ticked by a stationary atomic clock at $\vec\xi$ and it only coincides with $\tau$ at the origin, where $\vec\xi = 0$. 
In general, $\hat\tau \neq \tau$ and usually the readings of proper time $\hat\tau$ by two stationary clocks at two different places $\vec\xi_1 \neq \vec\xi_2\,$ will not keep synchronized. This will happen only if 
$$\left(\vec\xi_1 -\vec\xi_2\right)\cdot \vec a(\tau) = 0\,, \quad \forall \tau\in \mathbb{R}\,,$$ 
which only admits a solution if all directions $\vec a(\tau) = 0\,$ keep in the same plane.
It will be thus convenient to use the {\em synchronous time} $\tau\,$ instead of the local proper time $\hat\tau$.
 
The factor $\,1 + \vec\xi\cdot\vec{a}(\tau) \,$ is relevant in connexion with the domain of the FW coordinates, which does not embrace the whole Minkowski spacetime. Indeed, the procedure to obtain the coordinates of a point relies on solving the implicit function (\ref{E1}), which requires that the $\tau$-derivative of the left hand side does not vanish, that is $\,1 + \vec\xi\cdot\vec{a}(\tau) \neq 0\,$.

The unit velocity vector (with respect to a Lorentzian frame) of the worldline $\vec\xi =$constant in the FW reference frame at the synchronous time $\tau$ is
$$ u^\mu(\tau\,\vec\xi) = \frac{1}{\left| 1 + \vec\xi\cdot\vec{a}(\tau) \right|}\, \partial_\tau \varphi^\mu(\tau,\vec\xi) = u^\mu(\tau) \, \sign\left[1 + \vec\xi\cdot\vec{a}(\tau) \right] $$
and we shall restrict the domain of the FW coordinates to the region $\,1 + \vec\xi\cdot\vec{a}(\tau) > 0\,$ to avoid time reversal.
The hypersurface $\,1 + \vec\xi\cdot\vec{a}(\tau)  = 0\,$ or, in Lorentzian coordinates, $\,1 + a_\mu(\tau)\,\left[x^\mu - z^\mu(\tau)\right] = 0\,$ is the horizon of the Fermi-Walker coordinate system.

As for the proper acceleration of the worldline at $\vec\xi$, we have
\begin{equation}  \label{E6aa}
a^\mu(\tau;\vec\xi) = \frac{\D u^\mu}{\D\tau}\,\frac{\D\tau}{\D\hat\tau} = \frac{a^\mu(\tau)}{1 + \vec\xi\cdot\vec{a}(\tau)} \,, \qquad\qquad
a^j(\tau;\vec\xi) =  \frac{a^j(\tau)}{1 + \vec\xi\cdot\vec{a}(\tau) }\, 
\end{equation}
and the invariant proper acceleration $\,a(\tau,\vec\xi) = \sqrt{a^\mu(\tau,\vec\xi)\,a_\mu(\tau,\vec\xi)} \,$ is
$$  a(\tau,\vec\xi) = \frac{a(\tau)}{1 + \vec\xi\cdot\vec{a}(\tau)} $$
which differs from one place to another. 

It is worth to mention here that Einstein's statement \cite{Einstein1913b}: \guillemotleft 
... {\em acceleration} possesses as little absolute physical meaning as {\em velocity} \guillemotright, 
does not hold {\em avant la lettre}. As a matter of fact, every place $\vec\xi$ in a FW reference space has a proper acceleration (\ref{E6aa}) which is measurable with an accelerometer. However, the laws of classical particle dynamics also hold in the accelerated reference frame provided that a field of {\em inertial} force $-a^\mu(\tau;\vec\xi)$ is included; the passive charge for this field being the inertial mass of the particle. It is only with this specification that two accelerated reference frames are equivalent from the dynamical (or even physical) viewpoint.

Also notice that, as local proper acceleration is different from place to place, there is not such a thing as {\em the acceleration} of a FW reference frame and, when we use this expression, the origin acceleration should be understood. 

\subsection{ Uniqueness \label{S1.2}}
We shall now see that the form of the invariant interval (\ref{E6}) is unique in Minkowski spacetime provided that there is a synchronous reference whose reference space is flat.

\begin{proposition}
If in a given coordinate system the Minkowski metric components are $\, g_{ij} = \delta_{ij} \,, \quad  g_{4j} = 0\,$, $\quad i, j = 1 \ldots 3 \,$,
then there exist three functions $a_j(\tau)$ such that $ g_{44} = - \left[b + \vec\xi\cdot\vec a(\tau)\right]^2\,$, with $\,b= 0,\, 1\,$.
\end{proposition}

Indeed, let $(\xi^j,t)$ be coordinates for such a reference frame, being $\xi^j$ cartesian coordinates for the flat reference space. As the frame is synchronous,  the invariant infinitesimal interval is 
$$ \D s^2 = \D \vec\xi^2 - e^{2\phi}\,\D t^2 \,, \qquad {\rm with} \qquad \phi(\xi^j,t)  $$

It easily follows that all Christoffel symbols vanish except
$$ \Gamma^i_{44} = e^{2\phi} \,\partial_i\phi\,, \qquad  \Gamma^4_{\nu 4} = \partial_\nu \phi  $$ 
and the only non-vanishing components of the curvature tensor are
$$ R^4_{\;\;i4j} = - \partial_{ij} \phi -\partial_i \phi \, \partial_j \phi = e^{-\phi} \partial_{ij} e^\phi $$
As in Minkowski spacetime the curvature tensor is null, we have that $\,e^\phi = B(t) + \vec\xi\cdot\vec a(t)\,$, for some functions $B(t)$ and $a_i(t)$. Now, if $B(t) \neq 0$ the time coordinate can be redefined so that $\,d\tau = B(t)\,\D t\,$, and the proposed result immediately follows. \hfill $\Box$

The following Proposition is a kind of converse result:

\begin{proposition}
If in some coordinate system $(\xi^j,\tau)$ the Minkowski spacetime invariant interval is (\ref{E6}), then $(\xi^j,\tau)$ are the Fermi-Walker coordinates based on some origin worldline.
\end{proposition} 

Indeed, take the three functions $a_i(\tau)$ in the time rate of the interval (\ref{E6}), then set up the matrix $W^\beta_{\;\;\alpha}$ as indicated in the linear ordinary differential system (\ref{E2b}) and solve it for some arbitrary initial data $\, e^\mu_{(\beta)}(0)\,$. Take then $u^\mu(\tau) =  e^\mu_{(4)}(\tau) \,$ as the proper velocity and obtain the worldline $\,z^\mu(\tau)\,$ by integration for some initial $z^\mu(0)$. 

It is obvious that the tetrad  $\, e^\mu_{(\beta)}(\tau)\,$ is Fermi-Walker transported (\ref{E2}) along $\,z^\mu(\tau)\,$ and the Fermi-Walker coordinate transformation (\ref{E6a}) connects the given coordinates $(\xi^j,\tau)$ with Lorentzian coordinates. As the above procedure allows for arbitrary choices concerning initial data the result is not unique: the origin worldline is determined apart form the initial position $z^\mu(0)$, and the initial values of the tetrad  $\, e^\mu_{(\alpha)}(0)\,$ can be arbitrarily chosen.
\hfill $\Box$

\section{Generalised isometries \label{S2} }
To derive explicit expressions for the transformations connecting two different FW coordinate systems would require to invert the transformation law (\ref{E6a}), which generally is not feasible in closed form, except if the accelerations of both frames are constant and parallel to each other. In such a case the problem would be essentially one-dimensional and we could choose the space axes in the standard configuration, that is mutually parallel, with $X$ and $X^\prime$ parallel to the accelerations; then we should proceed as in Section 3 of ref. \cite{Llosa2015}.

In the general case, $a_i(\tau)$  variable and arbitrary, we can only derive expressions for infinitesimal transformations and, to this end, the notion of {\em generalized isometry} advanced by Bel \cite{Bel93} is helpful.

In any FW frame the invariant interval has the generic form (\ref{E6})  
\begin{equation}  \label{E6r}
\D s^2 = \D\vec{\xi}^2 - \left[1 + \vec{\xi}\cdot\vec{a}(\tau) \right]^2 \D\tau^2
\end{equation}
where $a_i(\tau)$ are some arbitrary functions of one variable. Therefore, the transformation formulae $\,(\xi^i,\tau) \longleftrightarrow (\xi^{\prime j},\tau^\prime)\, $ connecting any two FW frames must preserve the form (\ref{E6r}), perhaps with two different triples of functions $\left(a_i(\tau) \right)$ and $\left(a^\prime_i (\tau^\prime)\right)\,$, i. e. it is not an isometry because the metric is not invariant but {\em almost invariant} because only the functions $a_i(\tau)$ change whereas the overall form is preserved. 

To find the infinitesimal generalized isometries ---or generalized Killing vectors--- we write the interval as 
$$ \D s^2 = g_{\alpha\beta}(x, a_i(x)) \D x^\alpha\,\D x^\beta \,, \qquad $$
$a_i(x)\,,\quad  i = 1 \ldots m\,$ being a number of arbitrary functions ---in our particular case $m=3$--- and consider the infinitesimal transformations
$$ x^{\prime \alpha} = x^\alpha + \varepsilon\,X^\alpha(x) \,, \qquad a^\prime_i(x) = a_i(x) + \varepsilon\,A_i(x) $$
Then, as 
$$ g_{\mu\nu}(x^\prime, a^\prime_i(x^\prime)) \D x^{\prime \mu}\,\D x^{\prime \nu} = g_{\alpha\beta}(x, a_i(x)) \D x^\alpha\,\D x^\beta \,, $$
and keeping only first order terms in $\,\varepsilon\,$, we have that
$$ X^\alpha \partial_\alpha g_{\mu \nu}\left( x, a_i(x) \right) + 2\,  \partial_{(\nu} X^\alpha\, g_{\mu)\alpha} \left( x, a_i(x) \right) + \sum_i A_i \left(\frac{\partial g_{\mu\nu}}{\partial a_i}\right)_{\left( x, a_j(x) \right)} = 0  $$
or,  equivalently, the generalized Killing equation \cite{Bel93}
\begin{equation}  \label{E17}
 X_{\mu\|\nu} + X_{\nu\|\mu} + \sum_i A_i \frac{\partial g_{\mu\nu}}{\partial a_i} = 0 \,,
\end{equation}
where ``$\|$'' means covariant derivative.

In the particular case of the interval (\ref{E6}), which contains only three function $a_i(\tau)$ depending on only one variable,  we have that the above equation reduces to
$$ X^\alpha \partial_\alpha g_{\mu \nu}\left(x, a_j(\tau) \right) + 2 \,  \partial_{(\nu} X^\alpha \, g_{\mu)\alpha} + \sum_{i=1}^3 A_i(\tau)\,
\frac{\partial  g_{\mu \nu}\left( x, a_j(\tau)\right)}{\partial a_i}  = 0 \,,$$
that is
\begin{eqnarray}   \label{E18a}
\mbox{space block,}  & \; & \partial_i X_j + \partial_j X_i = 0 \\[1ex]  \label{E18b}
\mbox{cross block,}\hspace*{.5em}  & \; & \partial_4 X_i - \left(1 + \vec\xi\cdot \vec a \right)^2\,   \partial_i X^4  = 0 \\[2ex]  \label{E18c}
\mbox{time block,}\hspace*{.5em}  &  \; & \left(1 + \vec\xi\cdot \vec a \right) \,\left\{ X^\nu \partial_\nu \left(1 + \vec\xi\cdot \vec a \right) +  
\left(1 + \vec\xi\cdot \vec a \right) \,\partial_4 X^4 + \vec\xi\cdot\vec A \right\}= 0 
\end{eqnarray}

The space block implies that the spatial dependence of the components $X^i = X_i$ is the same as for a Killing vector of flat Euclidean metric $\delta_{ij}$, that is:
\begin{equation}   \label{E19}
\vec X(\tau) =  \vec f(\tau) + \vec\xi \times \vec\omega(\tau)
\end{equation}
where the usual 3-dimensional vector notation has been adopted for brevity, and  $\vec f(\tau)$ and $\vec \omega (\tau)$ are arbitrary functions. 

On its turn, the cross block gives the expressions for the three spatial derivatives $\partial_i X^4$, which carry as integrability conditions that
\begin{equation}   \label{E19a}
 \dot{ \vec\omega} = - \vec a \times \dot{\vec f}
\end{equation}
where a ``dot'' means $\partial_\tau\,$. Substituting (\ref{E19}) in equation (\ref{E18b}) and including (\ref{E19a}), we arrive at
\begin{equation}  \label{E20}
X^4 =  \frac{\vec\xi \cdot \dot{\vec f}(\tau)}{1 + \vec\xi\cdot \vec a(\tau)  } + g(\tau)
\end{equation}
where $g(\tau)$ is an arbitrary function. 

Finally, introducing the expressions (\ref{E19}) and (\ref{E20}) in the time block and including the separation of space and time variables, we obtain the following conditions on $g$, $\vec f$ and $\vec \omega$:
\begin{eqnarray}   \label{E21a}
 & & \partial_\tau g + \vec f \cdot \vec a = 0  \\[2ex]  \label{E21b}
 & & \partial_\tau\,\left( \partial_\tau \vec f + g\,\vec a\right) + \vec\omega \times \vec a + \vec A = 0 
\end{eqnarray}

These equations are solved in Appendix A in terms of 4-dimensional variables, namely the 4-vector $f^\mu(\tau) = \left( \vec f,\,g \right)$ and the skewsymmetric tensor  $\Omega_{\mu\nu}(\tau)$ formed with $\vec F = \dot{\vec f} + g\,\vec a\,$ as electric part and $\,\vec\omega\,$ as magnetic part
---see equation (\ref{A6}). The solutions (\ref{A8}) and (\ref{A9}) depend on ten constant parameters,  $f_0^\mu$ and $\Omega^0_{\alpha\beta}\,$, plus three arbitrary one-variable functions, $A_i(t)$. Introducing then these solutions in the expressions (\ref{E19}) and (\ref{E20}), we have that the infinitesimal generator is
\begin{equation}  \label{E22}
\mathbf{X} = \left[  f^\mu(\tau) +\Omega^\mu_{\; \; j} (\tau) \xi^j  \right]\,\hat\partial_\mu + \int_\mathbb{R} \D t \,\sum_{i=1}^3 A_i(t)\, \frac{\delta\quad}{\delta a^i(t)} 
\end{equation}
where $\hat\partial_i$ is the partial derivative with respect to $\xi^i$ and  $\hat\partial_4 = \displaystyle{\frac{1}{ 1 + \vec\xi\cdot \vec a(\tau)} \,\partial_\tau }\,$, that is a kind of {\em normalized partial derivatives}.
This infinitesimal generator acts on a manifold coordinated by $\,(\xi^j,\tau,[a_i(t)])\,$, where $\,a_i \in \mathcal{C}^0(\mathbb{R})\,$ and $\,(\xi^j,\tau) \in \mathbb{R}^4\,$ satisfy the condition $\,1+ \vec\xi \cdot \vec a(\tau)> 0 \,$. As the generator depends on the constant parameters $f_0^\mu$ and $\Omega^0_{\alpha\beta}\,$ and three arbitrary functions $A_i(t)$, we can separate this dependence as
\begin{equation}  \label{E23}
\mathbf{X} = f_0^\mu\,\mathbf{P}_\mu + \frac12\, \Omega_0^{\alpha\beta}\,\mathbf{J}_{\alpha\beta} +  \int_\mathbb{R} \D t \, \sum_{i=1}^3 A_i(t)\, \mathbf{D}_{i (t)} 
\end{equation}
where:
\begin{eqnarray}  \label{E23a}
\mathbf{P}_\mu & = & \Lambda_\mu^{\;\;\nu}(\tau)\,\hat\partial_\nu  \,, \qquad \qquad 
\mathbf{J}_{\alpha\beta}= 2\,k_{[\alpha}\,\mathbf{P}_{\beta]}  \\[2ex]  \label{E23c}
\mathbf{D}_{i (t)}  & = &  \frac{\delta\quad}{\delta a^i(t)} + 2\,\chi(t,\tau)\, \Lambda^{\mu [4}(t)\,\Delta^{i]}\,(\tau,t,\vec\xi) \,\mathbf{P}_\mu 
\end{eqnarray}
with $\quad \chi(t,\tau)= \theta(t)\theta(\tau-t) - \theta(-t)\theta(t-\tau)\,$ and  
\begin{eqnarray} \label{E23d}
k^\beta(\tau,\vec\xi) & = & \,\Lambda^\beta_{\;\;j}(\tau)\,\xi^j + \int_0^\tau \D t^\prime\, \Lambda^\beta_{\;\;4}(t^\prime) 
\\[2ex]  \label{E23e}
\Delta^\nu \left(\tau,t,\vec\xi\right) &=& \xi^j\,G_j^{\;\;\nu}(\tau,t) + \int_t^\tau \D t^\prime\, G_4^{\;\;\nu}(t^\prime,t)
\end{eqnarray}
(The matrices $\,\Lambda_\mu^{\;\;\nu}(\tau)\,$ and $\,G^\mu_\nu(\tau,t)\,$ are defined in Appendix A.)

The derivation of the Lie brackets between pairs of commutators is simple and we shall omit the details. However, as it is tedious and intricated, we explicite some useful intermediate formulae in Appendix B. The commutation relations are:
\begin{equation}  \label{E24}
\left.
\begin{array}{lll}
\left[\mathbf{P}_\mu,\mathbf{P}_\nu \right] = 0\,, & \qquad \left[\mathbf{J}_{\alpha\beta},\mathbf{P}_\mu  \right] = 2 \,\eta_{\mu[\alpha} \mathbf{P}_{\beta]}  \,, &  \qquad \left[\mathbf{J}_{\alpha\beta},\mathbf{J}_{\mu\nu}  \right] = 2\,\eta_{\mu[\alpha} \mathbf{J}_{\beta]\nu} - 2\,\eta_{\nu[\alpha} \mathbf{J}_{\beta]\mu}   \\[2ex]
\left[\mathbf{D}_{i(t)},\mathbf{P}_\mu  \right] = 0 \,,  & \qquad   \left[\mathbf{D}_{i(t)},\mathbf{J}_{\alpha\beta} \right] = 0   \,,& \qquad  
 \left[\mathbf{D}_{i(t)},\mathbf{D}_{j(t^\prime)}  \right] =  0
\end{array}   \right\}
\end{equation}

Thus, the algebra of the infinitesimal transformations connecting Fermi-Walker coordinate systems is an abelian extension of Poincaré algebra.

\section{Conclusion and outlook}
We have introduced a class of reference frames with arbitrary accelerated motion, namely Fermi-Walker frames. Each one is determined by the worldline representing the frame's spatial origin and, as it can be easily checked, if the origin is in uniform rectilinear motion, the reference frame is Lorentzian.

Each Fermi-Walker frame is characterized by the coordinates $\tau$, $\xi^1$, $\xi^2$ and $\xi^3$, and also by the three components of the [proper] acceleration of the origin. These quantities are three functions of time which are measurable from inside the frame, i. e. without referring to anything external, by means of accelerometers. 

The transformations connecting the coordinates of any pair of frames in the Fermi-Walker class preserve the form (\ref{E6}) of the spacetime interval, maybe with different functions $a^i(\tau)$. Thus we refer to these transformations as {\em generalised isometries}. Infinitesimal generalised isometries satisfy the generalised Killing equation (\ref{E17}), whose solution is an infinite dimensional Lie algebra that contains Poincaré algebra and acceleration boosts as well. A close look at the commutation relations reveals that it is an Abelian extension of Poincaré algebra; hence it is rather trivial from a mathematical viewpoint. Nevertheless it is curious that, contrarily to Lorentz boosts, acceleration boosts commute with each other and with any other Poincaré generators. 

We must also remark that the notion of generalized isometry \cite{Bel93} permits to go beyond Kretschmann's idea that, since special relativity admits the widest isometry group, it contains the largest relativity postulate. Our approach here has led to an intermediate group, namely the group of generlised isometries of the interval (\ref{E6}), which is larger than Poincaré group but much smaller than the whole diffeomorfism group.

\section*{Acknowledgement}
The present work is supported by Ministerio de Economía y competitividad (Spanish Gov.) project nş FPA2013-44549-P

\section*{Appendix A}
We shall solve here the equations (\ref{E19a}), (\ref{E21a}) and (\ref{E21b}) for $g(\tau)$, $\vec f(\tau)$ and $\vec\omega(\tau)\,$. Introducing the new variable $\vec F = \dot{\vec f} + g\,\vec a\,$ these three equations amount to:
\begin{equation}   \label{A1}
\left.  \begin{array}{l}
        \dot{\vec f} + g\,\vec a = \vec F \\[2ex]
        \dot g + \vec f\cdot \vec a = 0
        \end{array}   \right\}
\qquad 
\left.  \begin{array}{l}
        \dot{\vec F} + \vec\omega \times \vec a = - \vec A(\tau) \\[2ex]
        \dot{\vec\omega}+ \vec a \times \vec F = 0
        \end{array}   \right\}
\end{equation}

Putting $f^\mu = \left( \vec f,\,g \right) $ and $F^\mu = \left( \vec F,\,0 \right) $, the first pair of equations can be written as
\begin{equation} \label{A2}
 \dot f^\mu + W^\mu_{\;\;\nu} f^\nu = F^\mu
\end{equation}
where $\,W^\mu_{\;\;\nu}\,$ is the matrix in (\ref{E2b}).

Consider now the solution $\, \Lambda^\mu_{\;\;\alpha}(\tau)\,$ of (\ref{E6b}). As the tetrad $\, e^\mu_{(\alpha)}(\tau)= \Lambda^\mu_{\;\;\alpha}(\tau)\,$ also satisfies equation (\ref{E2b}), we shall have that
\begin{equation}  \label{A3b}
 \dot\Lambda^\mu_{\;\;\nu} =  \Lambda^\mu_{\;\;\beta} W^\beta_{\;\;\nu}
\end{equation}
and, being $\Lambda^\mu_{\;\;\nu}$ a Lorentz matrix, its inverse $\, L^\mu_{\;\;\nu} = \Lambda_\mu^{\;\;\nu}$ is a solution of
\begin{equation}  \label{A3}
\dot L^\mu_{\;\;\nu} =  - W^\mu_{\;\;\beta} \,L^\beta_{\;\;\nu}  
\end{equation}
where the fact that  $\,W_\mu^{\;\;\beta} = - W^\beta_{\;\;\mu}\,$ has been included. (Indices are raised and lowered with $\eta_{\mu\nu} = {\rm diag}[1,1,1,-1]$.)

Hence, $\, L^\mu_{\;\;\nu}(\tau)\,C^\nu = C^\nu\,\Lambda_\nu^{\;\;\mu}(\tau)\,$, with $C^\nu$ constant, is a solution of the homogeneous part of equation (\ref{A2}). We can then solve the complete equation by the method of variation of constants, so obtaining:
\begin{equation}  \label{A4}
f^\mu(\tau) = f^\nu_0\,\Lambda_\nu^{\;\;\mu}(\tau) + \int_0^\tau \D t\, G^\mu_{\;\;\rho}(\tau,t) \, F^\rho(t) 
\end{equation}
where 
\begin{equation}  \label{A3a}
 G^\mu_{\;\;\rho}(\tau,t) = \Lambda_\nu^{\;\;\mu}(\tau)\,\Lambda^\nu_{\;\;\rho}(t)\,
\end{equation} 
acts as a kind of matrix Green function. 
This solves the first pair of equations (\ref{A1}) provided that $\vec F(\tau)$ is known.

It is worth noticing that, although the matrices $ \Lambda_\nu^{\;\;\mu}(\tau)\,, \quad \tau\in\mathbb{R}\,$ are not in general a one-parameter subgroup of Lorentz group (except in the case of one-directional motion, as commented above), the matrices $G^\mu_{\;\;\rho}(\tau,t)\,$ do have the group property:
\begin{equation}   \label{A4a}
G^\mu_{\;\;\nu} (\tau,t)\,G^\nu_{\;\;\lambda} (t,t^\prime) = G^\mu_{\;\;\lambda} (\tau,t^\prime)  
\end{equation}
and moreover:
$$ G^\mu_{\;\;\nu} (\tau,t) = G_\nu^{\;\;\mu} (t,\tau) \qquad {\rm and}  \qquad G^\mu_{\;\;\nu} (t,t)= \delta^\mu_{\nu}  $$

To solve the second pair of equations (\ref{A1}) we need the following result whose proof is straightforward.
\begin{lemma}   \label{L1}
Let $\Omega_{\mu\nu}^0\,$ be a skewsymmetric matrix  and $\Lambda^\mu_{\;\;\nu}(\tau) \,$ a solution of (\ref{A3b}). Then the matrix 
$$ \Omega_{\mu\nu}(\tau) = \Omega_{\alpha\beta}^0\,\Lambda^\alpha_{\;\;\mu}(\tau)\,\Lambda^\beta_{\;\;\nu}(\tau) $$ 
is skewsymmetric and satisfies 
\begin{equation}  \label{A5}
 \dot\Omega_{\mu\nu} = \Omega_{\rho\nu}\,W^\rho_{\;\;\mu} + \Omega_{\mu\rho}\,W^\rho_{\;\;\nu}
\end{equation}
\end{lemma}

Consider now the skewsymmetric matrix $\,\Omega_{\mu\nu}\,$ set up with $\vec\omega$ and $\vec F$ as magnetic and electric parts, respectively
\begin{equation}  \label{A6}
  \Omega_{ij} = \epsilon_{ijk}\omega^k \, , \qquad \qquad \Omega_{i4} = - \Omega_{4i} = F_i  
\end{equation}
In terms of it, the second pair of equations (\ref{A1}) becomes:
$$  \dot\Omega_{\mu\nu} = \Omega_{\rho\nu}\,W^\rho_{\;\;\mu} + \Omega_{\mu\rho}\,W^\rho_{\;\;\nu} + A_{\mu\nu}    $$
where $\,A_{\mu\nu}\,$ is a skewsymmetric matrix whose coefficients all vanish except $\,A_{i4} = - A_{4i} = -A_i(\tau)\,$.

Lemma \ref{L1} gives the general solution of the homogeneous part of the latter equation and the solution of the complete equation easily follows by the method of variations of constants, namely
\begin{equation}    \label{A8}
 \Omega_{\mu\nu}(\tau) =  \Omega_{\alpha\beta}^0\,\Lambda^\alpha_{\;\;\mu}(\tau)\,\Lambda^\beta_{\;\;\nu}(\tau)  + 2 \,\int_0^\tau \D t\,\sum_{i=1}^3 A_i(t)\,G_\mu^{\;\;[4}(\tau,t)\,G_\nu^{\;\;i]}(\tau,t) 
\end{equation}
where $G_\mu^{\;\;\rho}$ is obtained by raising/lowering the indices in the above matrix Green function (\ref{A3a}), and the definition of $A_{\mu\nu}$ has been included.

Finally, as $F^\mu(t) = \Omega^\mu_{\;\;4}(t)\,$, equation (\ref{A4}) leads to
\begin{equation}  \label{A9}
 f^\mu(\tau) = f_0^\nu \Lambda_\nu^{\;\;\mu}(\tau) + \Omega_{\alpha\beta}^0\,\int_0^\tau \D t \, \Lambda^{\alpha\mu}(\tau)\, \Lambda^\beta_{\;\;4}(t)  + 2\,\int_0^\tau \D t\, \int_0^t \D t^\prime \,\sum_{i=1}^3 A_i(t^\prime)\,  G^{\mu[4}(\tau,t^\prime)\,G_4^{\;\;i]}(t,t^\prime) 
\end{equation}
where (\ref{A4a}) has been  included.

\section*{Appendix B}
Although we do not give an explicit derivation of the commutation relations (\ref{E24}), we next list some formulae which are of great help to this task. From equations (\ref{E23d}) and (\ref{E23e}) it easily follows that:
\begin{equation}  \label{B1}
 \Delta^\nu \left(\tau,t,\vec\xi\right) = \left[ k^\rho(\tau,\vec\xi) -  \int_0^t \D t^\prime\, \Lambda^\rho_{\;\;4}(t^\prime)\right]\,\Lambda_\rho^{\;\;\nu} 
\end{equation}
and also:
$$ \Delta^j(t,t,\vec\xi) = \xi^j \qquad {\rm and} \qquad \Delta^4(t,t,\vec\xi) = 0 $$

The commutators between the normalized partial derivatives and the functional derivatives are
$$ \left[ \frac{\delta\qquad}{\delta a^i(t)},\, \hat\partial_\nu \right] = - \delta_\nu^4 \,\frac{\xi^i}{1 + \vec\xi\cdot \vec a(\tau)}\,\delta(\tau -t)\,\hat\partial_4 $$

It is also easy to see that:
$$ \frac{\delta W_\rho^{\;\,\nu}}{\delta a^i(t)} =  2\,\delta(\tau - t)\, \delta^\nu_{[i} \eta_{4]\rho} $$
On this basis we can obtain a  differential equation satisfied by the functional derivative of $\Lambda_\mu^{\;\;\nu}(\tau)$, namely
\begin{equation}  \label{B2}
\partial_\tau\,\left(\frac{\delta \Lambda_\mu^{\;\,\nu}(\tau)}{\delta a^i(t)} \right) = \left(\frac{\delta \Lambda_\mu^{\;\,\rho}(\tau)}{\delta a^i(t)} \right)\, W_\rho^{\;\,\nu} + 2\,\delta(\tau-t)\,\Lambda_{\mu[4}(\tau) \delta^\nu_{i]}
\end{equation}
and, as the initial condition is $\Lambda_\mu^{\;\,\nu}(0) = \delta_\mu^\nu\,$, the initial condition for the above differential system is that
the functional derivative of $\Lambda_\mu^{\;\;\nu}(0)$ vanishes. Whence it easily follows that
\begin{equation}  \label{B3}
\partial_\tau\,\left(\frac{\delta \Lambda_\mu^{\;\,\nu}(\tau)}{\delta a^i(t)} \right) = 2\,\chi(t,\tau)\,\lambda_{\mu[4}(t)\, G^\nu_{\;\;i]}(\tau,t)
\end{equation}

Besides, the 4-vector $k^\nu(\tau,\xi)$  defined in (\ref{E23a}) has the nice properties that
\begin{equation}  \label{B4}
\mathbf{P}_\mu k_\nu = \eta_{\mu\nu} \,, \qquad \qquad \mathbf{D}_{i(t)} k_\nu = 0 
\end{equation}

\end{document}